%% file: oleskiewicz2019c.tex
\documentclass[fleqn,usenatbib]{mnras}

\usepackage[T1]{fontenc}
\usepackage{ae,aecompl}

\usepackage[a4paper]{geometry}
\usepackage[utf8]{inputenc}
\usepackage{booktabs}
\usepackage{siunitx}
\usepackage{graphicx}
\usepackage{pgf}
\usepackage{amsmath}
\usepackage{amssymb}
\usepackage[capitalise]{cleveref}
\usepackage{hyperref}
\usepackage{url}
\usepackage{xcolor}

\crefformat{equation}{Eq~(#2#1#3)}

\DeclareSIUnit\erg{erg}
\DeclareSIUnit\h{\textit{h}}
\DeclareSIUnit\Msolar{M_{\odot}}
\DeclareSIUnit\parsec{pc}
\DeclareSIUnit\yr{yr}

\newcommand{\GF}{\texttt{GALFORM} }
\DeclareMathOperator{\A}{\mathbf{A}}
\DeclareMathOperator{\B}{\mathbf{B}}
\DeclareMathOperator{\AB}{\mathbf{A}_\mathbf{B}^{\left(i\right)}}
\DeclareMathOperator{\Xwoi}{\mathbf{X}_{\sim{i}}}
\DeclareMathOperator{\Var}{\mathrm{Var}}

\title[GALFORM Sensitivity Analysis]{
Sensitivity analysis of a galaxy formation model}

\author[P. Ole\'{s}kiewicz et al.]{
	Piotr Ole\'{s}kiewicz$^{1}$
	\thanks{E-mail: piotr.oleskiewicz@durham.ac.uk},
	Carlton M. Baugh$^{1,2}$,
	\\
	$^{1}$Institute for Computational Cosmology, Department of Physics,
	Durham University, South Road, Durham DH1 3LE, UK.\\
	$^{2}$Institute for Data Science, Department of Physics,
	Durham University, South Road, Durham DH1 3LE, UK.
}

\date{Accepted XXX. Received YYY. in original form ZZZ}
\pubyear{2019}

\begin{document}
\label{firstpage}
\pagerange{\pageref{firstpage}--\pageref{lastpage}}
\maketitle

\begin{abstract}

We present the first application of a variance-based sensitivity analysis (SA)
to a model that aims to predict the evolution and properties of the whole galaxy
population. SA is a well-established technique in other quantitative sciences,
but is a relatively novel tool for the evaluation of astrophysical models. We
perform a multi-parameter exploration of the \GF semi-analytic galaxy formation
model, to compute how sensitive the present-day $K$-band luminosity function is
to varying different model parameters. The parameter space is scanned using a
low-discrepancy sampling technique proposed by Saltelli. We first demonstrate
the usefulness of the SA approach by varying just two model parameters, one
which controls supernova feedback and the other the heating of gas by AGN. The
SA analysis matches our physical intuition regarding how these parameters affect
the predictions for different parts of the galaxy luminosity function. We then
use SA to compute Sobol' sensitivity indices varying seven model parameters,
connecting the variance in the model output to the variance in the input
parameters.  The sensitivity is computed in luminosity bins, allowing us to
probe the origin of the model predictions in detail.  We discover that the SA
correctly identifies the least- and most important parameters. Moreover, the SA
also captures the combined responses of varying multiple parameters at the same
time. Our study marks a much needed step away from the traditional
``one-at-a-time" parameter variation often used in this area and improves the
transparency of multi-parameter models of galaxy formation.

\end{abstract}


\input{body.tex}

\section*{Acknowledgements}

This work was supported by the Science and Technology facilities Council
ST/L00075X/1.  PO acknowledges an STFC studentship funded by STFC grant
ST/N50404X/1.  The inspiration for this project came from a placement by PO at
Atom Bank, which was supported by the STFC/Durham University Centre for Doctoral
Training in Data Intensive Science, supported by ST/P006744/1.  This work used
the DiRAC Data Centric system at Durham University, operated by the Institute
for Computational Cosmology on behalf of the STFC DiRAC HPC Facility
(\url{www.dirac.ac.uk}).  This equipment was funded by BIS National
E-infrastructure capital grant ST/K00042X/1, STFC capital grants ST/H008519/1
and ST/K00087X/1, STFC DiRAC Operations grant ST/K003267/1 and Durham
University. DiRAC is part of the National E-Infrastructure.

\bibliographystyle{mnras}
\bibliography{oleskiewicz2019c}

\bsp
\label{lastpage}
\end{document}

%% file: body.tex
\newcommand{\cB}[1]{\textcolor{blue}{#1}}
\newcommand{\cR}[1]{\textcolor{red}{#1}}

\section{Introduction}\label{sec:3_intro}

Galaxy formation is a complex process which we are only now just starting to
understand through a combination of observations, numerical simulations and
analytical modelling. Two main theoretical techniques are used to model the
formation and evolution of galaxies: semi-analytical modelling (SAM) and
hydrodynamic simulations (for a review see \citealt{somerville2015}).  SAMs use
physically motivated, simplified mathematical relations to describe the
evolution of baryons in growing dark matter haloes \citep{baugh2006,
benson2010b}. Hydrodynamic simulations, on the other hand, tend to make fewer
assumptions and approximations than SAMs and solve the fluid equations governing
the dynamics of baryons. Nevertheless, in hydrodynamic simulations many
processes, such as star formation, remain ``sub-grid'' due to the finite
numerical resolution of the simulation and our inability to write down the
precise equations describing some processes \citep{crain2015,ludlow2019}. In the
absence of a complete mathematical description, physical processes are described
in both SAMs and hydrodynamic simulations by approximate equations that contain
parameters.  Values have to be chosen for these parameters to specify a model.
Here, we present a new application of an established statistical method to
assess the impact of changes in model parameters on the output of a model.   

The past few years have seen tremendous breakthroughs in the hydrodynamic
simulation of galaxy formation for significant galaxy populations in
cosmological volumes \citep{vogelsberger2014,schaye2015,pillepich2018}.
Nevertheless, SAMs remain an attractive and valuable complement to
hydrodynamical simulations due to their flexibility and speed. These properties
of SAMs mean that they can be used to build intuition about physical processes,
by running thorough investigations of the impact of varying model parameters
(e.g. see the comprehensive exploration of perturbations around the fiducial
model presented by \citealt{lacey2015}). Also, SAMs remain the method of choice
to populate large volume N-body simulations using a physical galaxy formation
model: the fiducial simulation volumes used in SAMs are around 100 times bigger
than those used in the current state-of-the-art hydrodynamical simulations. The
predictions of SAMs have reached an impressive level of maturity through careful
comparisons between the predictions of different groups and techniques (e.g.
\citet{contreras2013,knebe2015,guo2016,mitchell2018}).  

Nevertheless, some scepticism remains regarding SAMs, much of which can be
traced to the way in which the model parameters are set. Traditionally models
have been calibrated by developing physical intuition about how the model
responds to changes in selected parameter values, such as those which control
the mass loading of winds driven by supernovae, and then varying one parameter
at a time to hone in on a best-fitting model. Often the quality of the model
reproduction of the calibration data is judged by eye and compromises are made
in order to match multiple datasets; these steps are hard to quantify and
therefore difficult to reproduce. The ``best-fitting'' model is reported as a
single choice of parameter set that defines the model. The primary motivation
for producing a single model is the desire to build mock catalogues for galaxy
surveys \citep{baugh2013}. However, users often want to know the uncertainty on
the model predictions and how the predictions respond to changes in the input
parameters.

The range of processes modelled by SAMs lends them the flexibility to predict
varied observation but at the cost of having to specify a number of parameters
which complicate model optimisation or calibration. A number of techniques have
been devised to reduce the complexity or dimensionality of the parameter space
and to perform efficient searchs of the parameter space: principal component
analysis \citep[hereafter PCA]{benson2010a}, Bayesian emulators
\citep{bower2010,gomez2012}, particle swarm optimiser \citep[hereafter
PSO]{ruiz2015}, Markov Chain Monte Carlo \citep{henriques2009, lu2011, lu2012,
henriques2013, mutch2013, martindale2017}, and Latin-hypercube sampling
\citep{bower2010, rodrigues2017}.  

Here we apply sensitivity analysis to quantify the dependence of the model
output on the variation in the values of the model input parameters. The
analysis of \citet{gomez2014} using the ChemTreeN SAM of \citet{tumlinson2009}
is similar in scope to our work.  They use an analysis of variance technique for
variance decomposition instead of sensitivity indices, and Gaussian processes
for model fitting. Here we use the \GF SAM effectively as a black-box model, and
evaluate the sensitivity of the model outputs to the variation of the input
parameters. A SAM is an ideal candidate for sensitivity analysis, as the
interactions between parameters are complex enough to develop a black-box-like
behaviour (``becomes easier to experiment with than to understand'')
\citep{golovin2017}; however, many parameters have a natural physical
interpretation, and hence it will be straightforward to develop intuition about
how sensitive the model outputs should be to changing the  inputs.  Many
parameters also have either physically motivated bounds, or at least a plausible
range of possible values.

A criticism often aimed at SAMs is that they contain too many free parameters.
This is usually rebuffed with the insistence that the parameters are physical,
not statistical. Model fitting alone is therefore insufficient for interpreting
how well a SAM is performing.  A different research question, one this study
tries to address, is how sensitive the model is to the parameter variation -- in
other words, how well do we understand the impact of the physical processes and
their interactions on the model predictions?

Sensitivity analysis (SA) \citep{fisher1918,sobol1993,sobol2001,saltelli2010} is
an area of statistical modelling which analyses how the variance of the output
of a model is affected by variance in the model inputs.  It is closely related
to uncertainty analysis and model optimisation, and can be used to test the
robustness of the model predictions to uncertainty in the input parameters,
quantify dependence of the outputs of a model on different parameters, identify
model non-linearities, and guide subsequent model optimisation.  This addresses
a common criticism of black-box models, namely that after adding sufficiently
many free parameters they can be fine tuned to match any observations, and
provide a single set of predictions.  While model optimisation can be used to
compute confidence intervals, SA is uniquely positioned to quantify model
responses and the relative importance of the inputs.  This addresses the
complaint about SAMs listed above, that providing a spread of model predictions
is preferable to fitting to the observations.  Using SA, we will be able to not
only tell how much model predictions vary for individual outputs, but also
quantify how much of this variance can be attributed to individual model inputs
(or their combinations).

There are several SA techniques, not all of which are suitable for analysing
non-linear models with a  high-dimensional parameter space.  With a few
exceptions\footnote{Some methods, such as Gaussian processes, use parameter
exploration to simultaneously measure model sensitivity and maximise
goodness-of-fit for model output(s).}, SA is done in 3 stages:
\begin{enumerate}
	\item sampling of the parameter space
	\item model evaluation in the parameter space 
	\item computation of sensitivity indices
\end{enumerate}

Here, we use a variance-based SA which adopts the improvement of introduced by
\cite{saltelli2017} over the  Sobol' indices.  Variance-based methods aim to
decompose the variance of the model output into the contributions from
individual parameter variances, as well as the combined variances of the
interactions of multiple combinations of parameters changing at once.  In order
to avoid a computational penalty for evaluating all possible parameter
combinations, input parameters are treated as probability distributions, and the
sensitivity of the model output is estimated approximately.  Moreover, a number
of numerical optimisations have been introduced into the sampling and index
calculation techniques, to improve the convergence of the indices and average
over the values which are too difficult to compute efficiently.

This work diverges from previous studies in two important ways: firstly, we
narrow the scope of this investigation to computing only sensitivity indices,
and we do not attempt to provide the best-fitting values for a galaxy formation
model.  We believe that SA is not the best tool for this task, as it
investigates model responses at the extreme values of input parameters, and
often for unusual combinations of inputs, where the model no longer reproduces
the observable values.  Secondly, we do not limit ourselves to measuring
responses of the model to individual parameters and their linear combinations.
Instead, we use sensitivity indices to capture both individual and combined
impacts of parameters.  Lastly, this study focuses exclusively on one
observable, the $K$-band luminosity function, calculated using the \GF SAM, and
probes how this specific model reacts to changes in the input parameters.  Our
scope is narrower, but also deeper than any previous study in this area.

The layout of the paper is as follows. In Section~2 we set out the theoretical
background, introducing the \GF model and, for completeness, giving the
equations for the processes that we vary (\S~2.1). We then discuss variance
based sensitivity analysis (\S~2.2), the concept of low-discrepancy sampling
(\S~2.3), the exploration of parameter space using Saltelli sampling (\S~2.4),
define the sensitivity indices (\S~2.5) and illustrate these ideas with a toy
model (\S~2.6). Our results using \GF are presented in Section~3 and our
conclusions are given in Section~4. 

\section{Theoretical background}\label{sec:3_theory}

Here we set out the theoretical ideas used in the paper. \S~2.1 gives a brief
overview of the \GF semi-analytical model, introducing the processes that are
varied in the sensitivity analysis. \S~2.2 introduces variance based sensitivity
analysis, \S~2.3 discusses the sampling of a model parameter space and \S~2.4
covers Saltelli sampling. \S~2.5 defines the sensitivity indices and \S~2.6
illustrates their use with a toy model. \S~2.7 discusses the use of \GF output
in the sensitivity analysis. 

\subsection{GALFORM}\label{sec:3_galform}

\begin{table}
	\centering
	\caption{\protect\citet{planckcollaboration2014} cosmology used in the
	P-Millennium simulation; the last two rows give the simulation box length
	and the number of particles used.}
	\label{tbl:3_planck}
	\begin{tabular}
		{lr}
		\toprule
		parameter               & value    \\
		\midrule
		$\Omega_{\Lambda}$      & 0.693    \\
		$\Omega_M$              & 0.307    \\
		$\Omega_{\rm baryon}$   & 0.04825  \\
		$h$                     & 0.6777   \\
		$\sigma_8$              & 0.8288   \\
		$n$                     & 0.967    \\
		$L[h^{-1}\mathrm{Mpc}]$ & 542.16   \\
		$N_{\mathrm{P}}$        & $5040^3$ \\
		\bottomrule
	\end{tabular}
\end{table}

\GF is a SAM which aims to predict the properties of galaxies starting from dark
matter halo merger histories that are typically extracted from an N-body
simulation \citep{cole2000,baugh2006,bower2006,lacey2015}. \GF models the
processes which shape the galaxy population using a set of physically motivated,
non-linear differential equations which track the exchange of mass, energy and
angular momentum between the different components of a galaxy.  The processes
modelled are:
\begin{enumerate}
	\item the merger histories of dark matter halos
	\item the heating and cooling of gas and the formation of galactic discs 
	\item quiescent star formation in galactic discs
	\item bursts of star formation triggered by galaxy mergers or dynamically unstable disks
	\item feedback driven by supernovae (SNe), which can eject cold gas from a galaxy
	\item heating by an active galactic nucleus (AGN), which can prevent gas cooling
    \item chemical enrichment of stars and gas  
\end{enumerate}

These processes are in many cases modelled by equations that contain parameters.
A \GF model corresponds to a set of parameters whose values have  been chosen so
that the model reproduces a particular set of observations. Some of these
parameters govern different choices for processes in the model, such as the
radial density profile assumed for the hot gas within a halo or the stellar
initial mass function (IMF) which describes the number of stars of different
masses produced in episodes of star formation. For example, the
\cite{gonzalez-perez2014} model assumes a universal, solar neighbourhood IMF
whereas the \cite{lacey2015} model invokes a top-heavy IMF in bursts of star
formation and a solar neighbourhood IMF in quiescent star formation. Even though
these two models are implemented in the same N-body simulation, the choices made
regarding the IMF and the slightly different emphasis on which observations the
model should reproduce most closely means that there are several differences in
the values of the parameters which define these galaxy formation models. 

Here we use the recalibration of the \cite{gonzalez-perez2014} model introduced
by \citet{baugh2019} for the  Planck Millennium $N$-body simulation, which we
refer to as GP14.PMILL. The Planck Millennium N-body simulation (hereafter the
PMILL simulation) adopts the Planck cosmology
(\citealt{planckcollaboration2014}; see Table~1) and has superior mass
resolution and halo merger histories that are better sampled in time compared
with earlier N-body simulations into which \GF was implemented (see Table~1).
Below we review the processes that we vary in the sensitivity analysis. A full
description of \GF can be found in \cite{lacey2015}.

\subsubsection{Star formation rate}\label{sec:3_sfr}

The GP14.PMILL model uses an empirically motivated star formation law that was
introduced by \citet{blitz2006} and implemented in \GF by \cite{lagos2011}. The
star formation rate is given by 
\begin{equation}\label{eqn:3_alpha_ret}
\Sigma_{\mathrm{SFR}}=\nu_{\mathrm{SF}}\times{f}_{\mathrm{mol}}\times\Sigma_{\mathrm{gas}},
\end{equation}
where $\Sigma_{\mathrm{SFR}}$ is the star formation rate per unit area,
$\Sigma_{\mathrm{gas}}$ is the surface density of gas, $\nu_{\mathrm{SF}}$ is
the inverse of the star formation time-scale, and $f_{\mathrm{mol}}$ is the
ratio of the surface densities of the molecular and total gas masses,
$\Sigma_{\mathrm{mol}}/\Sigma_{\mathrm{gas}}$.

\subsubsection{Supernova feedback}\label{sec:3_sn}

Supernova feedback in \GF is modelled as a process which ejects cold gas from a galaxy to a reservoir of mass $m_{\mathrm{res}}$, at a rate of
\begin{equation}\label{eqn:3_m_out}
\dot{m}_{\mathrm{out}}=\beta \psi,
\end{equation}
where $\psi$ is the star formation rate and $\beta$ is a mass loading factor defined as
\begin{equation}\label{eqn:3_beta}
\beta =\left(\frac{V_{\mathrm{c}}}{V_{SN}}\right)^{-\gamma_{\mathrm{SN}}}.\\ 
\end{equation}
Here $V_{\rm SN}$ and $\gamma_{\mathrm{SN}}$ are model parameters and $V_{\mathrm{c}}$ is the effective circular velocity of the disk or bulge (for starbursts)  at the half mass radius. Note that these equations are applied to quiescent and burst star formation. Different values can be adopted for the parameter $V_{\rm SN}$ for the disk and burst contributions to star formation. 

Gas is returned from this reservoir to the hot gas halo at the rate of
\begin{equation}\label{eqn:3_alpha_ret2}
\dot{m}_{\mathrm{ret}}=\alpha_{\mathrm{ret}}\times\frac{m_{\mathrm{res}}}{\tau_{\mathrm{dyn}}},
\end{equation}
which is controlled by the free parameter $\alpha_{\mathrm{ret}}$;
$\tau_{\mathrm{dyn}}=r_{\mathrm{vir}}/V_{\mathrm{vir}}$ is the dynamical time of
the halo, where $r_{\rm vir}$ is the virial radius of the halo and $V_{\rm vir}$
is the effective circular velocity at this radius.

\subsubsection{AGN feedback}\label{sec:3_agn}

Supermassive black holes (SMBHs) grow in the centres of galaxies, and inject
energy into the gas reservoir of a halo following accretion, which disrupts the
cooling process (see \citealt{fanidakis2011} and \citealt{griffin2019} for
descriptions of the treatment of SMBHs in \GF).  In \GF AGN heating occurs when
two conditions are met: (i) the hot gas halo is in quasi-hydrostatic
equilibrium, defined in terms of the ratio of the cooling time,
$\tau_{\mathrm{cool}}$, to the free-fall time, $\tau_{\mathrm{ff}}$:
\begin{equation}\label{eqn:3_alpha_cool}
\frac{\tau_{\mathrm{cool}}(r_{\mathrm{cool}})}{\tau_{\mathrm{ff}}(r_{\mathrm{cool}})}>\frac{1}{\alpha_{\mathrm{cool}}},
\end{equation}
where $\alpha_{\mathrm{cool}}$ is a parameter, and (ii) the AGN power required
to balance the radiative cooling luminosity $L_{\rm cool}$ is below a fraction
$f_{\rm Edd}$ of the Eddington luminosity $L_{\rm Edd}$ of the SMBH of mass
$M_{\rm BH}$: 

\begin{equation}
L_{\rm cool}  < f_{\rm Edd} L_{\rm Edd} \left(M_{\rm BH}\right).
\end{equation}

\subsubsection{Disc instabilities}\label{sec:3_stab}

Galaxies can also undergo morphological transformations and starbursts as a
result of disc instabilities. Galaxy discs which are dominated by rotational
motions are unstable to bar formation when they are sufficiently
self-gravitating. We assume that discs are dynamically unstable to bar formation
if \citep{efstathiou1982}:
\begin{equation}\label{eqn:3_f_stab}
F_{\rm disc} =  \frac{v_c(r_{\mathrm{disc}})}{\sqrt{1.68 \, G \, M_{\mathrm{disc}}/r_{\mathrm{disc}}}}<f_{\mathrm{stab}}, 
\end{equation}
where $M_{\rm disc}$ is the total disc mass (ie stars plus cold gas), $r_{\rm
disc}$ is the disc half-mass radius, and the factor 1.68 relates this to the
exponential scale length of the disc.

The quantity $F_{\rm disc}$  measures the contribution of disc self-gravity to
its circular velocity, with larger values corresponding to less self-gravity and
so greater disc stability.  \cite{efstathiou1982} found a stability threshold
$F_{\rm stab} \approx 1.1$ for a family of exponential stellar disc models.
Note that a completely self-gravitating stellar disc would have $F_{\rm disc} =
0.61 $, which is therefore the minimum value allowed for this parameter.

\subsubsection{Parameter selection}\label{sec:3_param}

\begin{table}
	\centering
	\caption{The \GF parameters analysed in this work.  The parameter ranges have been taken from previous analyses
	\protect\citep{bower2010,rodrigues2017}.}
	\label{tbl:3_param}
	\begin{tabular}
		{llrr}
		\toprule
		process            & parameter                                   & min  & max \\
		\midrule
		star formation     & $\nu_{\mathrm{SF}}$ [\si{\per\giga\yr}]     & 0.2  & 1.2 \\
		supernova feedback & $\gamma_{\mathrm{SN}}$                      & 1.0  & 4.0 \\
		                   & $\alpha_{\mathrm{ret}}$                     & 0.2  & 1.2 \\
		                   & $V_{\mathrm{hot,disk}}$ [\si{\km\per\sec}]  & 100  & 550 \\
		                   & $V_{\mathrm{hot,burst}}$ [\si{\km\per\sec}] & 100  & 550 \\

		AGN feedback       & $\alpha_{\mathrm{cool}}$                    & 0.2  & 1.2 \\
		disc instabilities & $f_{\mathrm{stab}}$                         & 0.61 & 1.1 \\
		\bottomrule
	\end{tabular}
\end{table}

We consider the relative importance of the processes described in \S~2.1.1 -
\S~2.1.4 by performing a SA on the parameters that describe these phenomena. The
parameters and the ranges over which they are varied are listed in Table~2. In
some instances, the parameter range is reasonably well defined, such as $f_{\rm
stab}$, as discussed above in \S~2.1.4. In other cases, the choice of range of
parameter values is less well defined. For example, using simple conservation
arguments, $\gamma_{\rm SN}$ could take on values of $1$ and $2$ in the momentum
and energy conserving phases of the wind evolution
\citep{ostriker1988,lagos2013}. Numerical simulations of winds have suggested
different values of $\gamma_{\rm SN}$. The other parameters defining the \GF
model beyond those listed in Table~2 are held fixed. 

\subsubsection{Model output}\label{sec:3_lf}

After the formation and evolution of galaxies is calculated over the merger
history of the dark matter haloes in the PMILL simulation, galaxy luminosities
can be obtained from the predicted star formation rate and metallicity of the
stars produced using a stellar population synthesis model.  Dust extinction is
calculated in post-processing, based on the size and gas metallicity of each
galaxy \citep{gonzalez-perez2014, lacey2015}. The model output that we focus on
here is the $K$-band luminosity function at $z=0$.

\begin{figure*}
\input{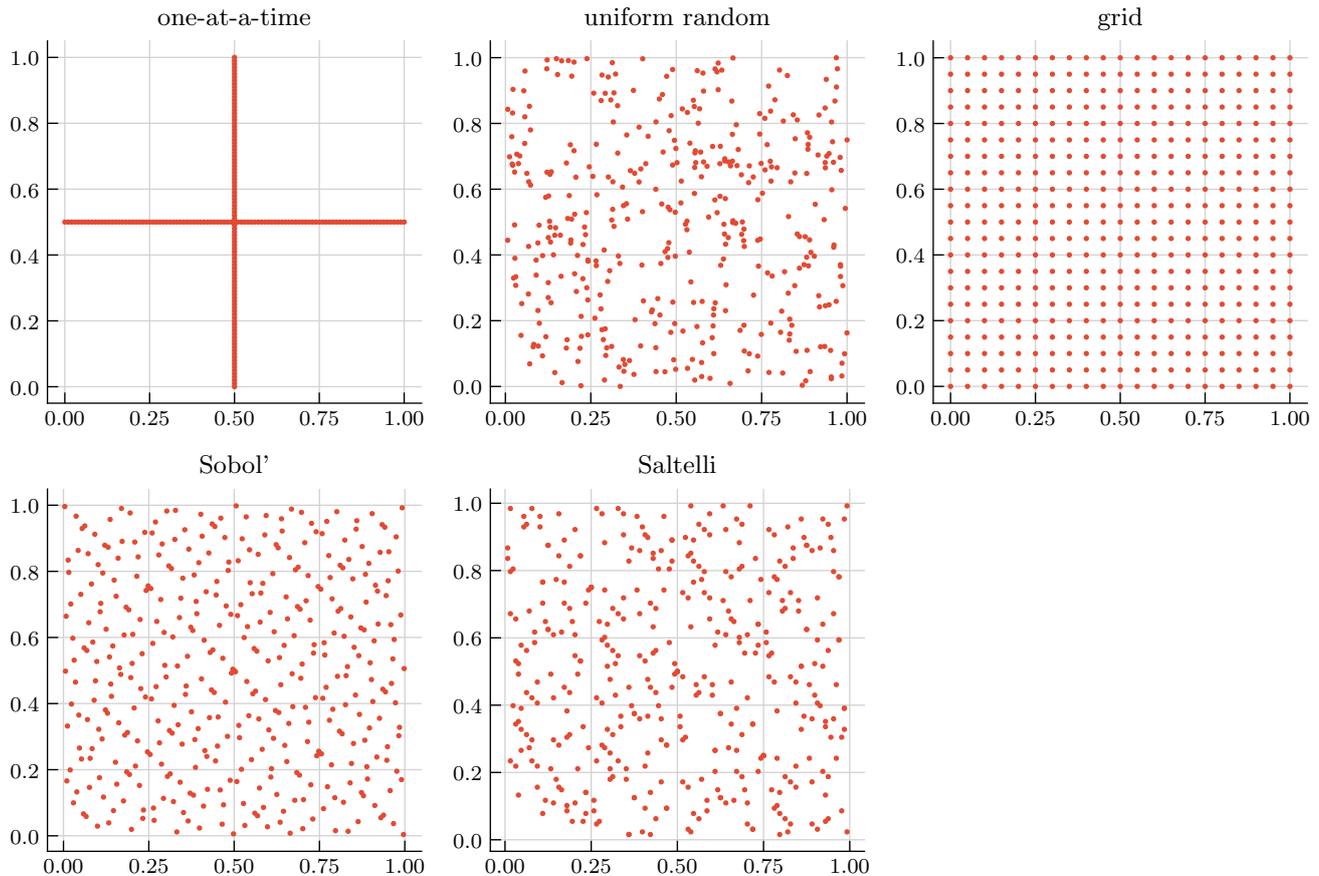}
\caption[Comparison of parameter space sampling strategies]{Comparison of
selected  parameter space sampling strategies.  Each panel contains 400 points
sampled between \([0,1]\) in the \(X_0\) and \(X_1\) dimensions using different
methods, as labelled in each panel.}
\label{fig:3_sample}
\end{figure*}

\subsection{Variance-based sensitivity analysis}\label{sec:3_sens}

The SA method we use here closely follows those used by \cite{sobol2001}  and
\cite{saltelli2017}, which are designed to decompose variance in the model
output into the variances of the input parameters and their interactions using
as few model evaluations as possible.

Many SA approaches suffer from a number of shortcomings which make them
unsuitable for analysing non-linear models. By non-linear models we mean here
ones that are characterised by interactions between the
inputs\footnote{Interactions between inputs occur, for example, when varying two
or more input parameters produces a significantly different response from the
model than would be expected from summing the change produced by varying the
parameters independently.} and which therefore cannot be analysed effectively
using regression or one-at-a-time (OAT) parameter variation techniques
\citep{morris1991}.

Unlike other methods, variance-based SA allows a full exploration of the input
space, and therefore accounts for the interactions between parameters and
non-linear responses of the model.  It follows that variance-based methods are
able to evaluate the total effect indices (see below) and rank the parameters in
order of their influence on the output \citep{chan1997, sobol2001,saltelli2010}.

Finally, we note that all SA methods assume that the model inputs are
independent, which might not hold in general for complex models.  For instance,
correlations between inputs, or unphysical combinations of their values, cannot
be recognised by SA techniques.  Similarly, variance-based SA currently assumes
that the model output is a scalar. This means that the model outputs are
independent of one another; for example in the case of the luminosity function,
the model prediction in a luminosity bin is considered to be independent of the
results in other bins and from other outputs, e.g. other galaxy statistics.
Even if the output of the model is multi-dimensional, and even if it is
correlated across one or more dimensions\footnote{We know that this is the case
in galaxy formation models because if the luminosity function changes in a given
bin this will lead, for example, to a change in the luminosity - circular
velocity relation. \cite{benson2014} argued that correlations between bins in
the observed luminosity function are important in setting model parameters.},
each of the outputs must be analysed in isolation from the others.
Unfortunately, at the time of writing there are {\it no} well-established
techniques which quantify or alleviate these two shortcomings.  However, these
limitations do not apply to \GF: the input parameters can all be varied
independently and freely across the entire parameter space, and our outputs will
be quantised and analysed independently.

\subsection{Sampling parameter space}\label{sec:3_sobol}

Sampling the high-dimensional parameter space of a complex model requires a
trade-off between the accuracy of the sampling and computational expense. The
accuracy of the sampling describes how well the space is probed -- have any
potentially interesting regions of the parameter space been overlooked because
too few points have been sampled or because the method used has left gaps in the
space? 

The accuracy of a sampling scheme can be assessed formally in terms of its
``discrepancy''. The lowest discrepancy sampling possible is a regular grid.
However, this is subject to ``aliasing'' or a lack of resolution due to the
fixed gaps in the parameter space between the model evaluations; interesting
model behaviour could be hidden in the unsampled parts of the parameter space.
The convergence of the exploration of the parameter space is slow with a regular
grid. The aliasing can be reduced and the convergence rate sped up by using a
random sequence to sample the parameter space, which leads to a higher density
of sampling in some parts of parameter space compared to a regular grid. The
drawback in this case is that some regions of the parameter space will be more
sparsely sampled than they were using a regular grid. A random sequence is
formally described as the highest discrepancy sampling. Ideally, for a fixed
number of sampling points, we want to strike a balance between avoiding the
regular sampling achieved using a grid and leaving big gaps unsampled in the
parameter space, as happens  with random sampling. 

Several quasi-random techniques have been proposed to generate sequences that
approach  this ideal of ``low discrepancy'' sampling, and which also ensure fast
convergence of the uncertainties in the sensitivity indices. A quasi-random
sequence is one designed to generate points in $d$-dimensions which appear
random, but which are generated deterministically to have certain desired
properties.  Unlike pseudo- and truly-random sequences, successive points in a
quasi-random sequence fill the gaps left by the previous points in the parameter
space.  The ``random'' part of the name is technically a misnomer, as the
sequence is fully deterministic, but yields a uniform distribution when
projected onto any dimension of the parameter space.

A quasi-random sequence can be designed to minimise its discrepancy.  For a
low-discrepancy sequence, all of its subsequences also have low discrepancy.  If
a given sequence is uniform, its discrepancy tends to zero as its length
increases.  For these reasons, quasi-random low-discrepancy sequences are used
to maintain a balance between rapid  convergence of numerical algorithms, a
thorough coverage of the parameter space, and a high uniformity of a resulting
sample along all dimensions of the parameter space \citep[][\S 7.8]{press2007}.
Quasi-random sequences are therefore an attractive replacement for pseudo-random
sequences in many applications which require a high quality sampling.

Sampling based on low discrepancy sequences, such as the recurrent additive
sequence \citep{ulam1960}, Halton sequence \citep{halton1964}, Latin hypercube
\citep{stein1987} or Sobol' sequence \citep{sobol1967, levitan1988} can be used
in numerical integration and model optimisation and have been shown to
outperform schemes based on truly random, or pseudo-random number generators,
while achieving significantly faster convergence rates \citep{sobol1993}.  The
advantage of these sequences over truly random and pseudeo-random sequences can
be attributed to the fact that the low discrepancy property guarantees gap-less
sampling over the entire parameter space. 

The low discrepancy quasi-random sequence typically used in SA is the Sobol'
sequence \citep{sobol1967}.  It can be efficiently calculated, and produces a
sample which quickly converges to the correct set of sensitivity indices, as
verified by checking against analytically calculated values for test models.
Even though it is impossible to estimate the required number of model
evaluations prior to running the SA, there exists a natural convergence
criterion -- the sum of the first-order indices, defined in \cref{eqn:3_s_i},
has to add up to unity. Moreover, even if the SA did not converge after the
initial run, additional evaluations can be easily added (see the example in the
next subsection).

\subsection{Saltelli sequence sampling}\label{sec:3_saltelli}

The Sobol' sequence was originally proposed as a method of improving the
convergence of numerical integration \citep{sobol1967}. \citet{antonov1979}
developed an efficient computational method to implement Sobol' sampling.
\citet{saltelli2010} combined multiple Sobol' sequences to further reduce the
number of points required for the estimation of the sensitivity indices,
improving the convergence rate.

Hereafter we refer to the Sobol' sequence as an $N$ by $d$ matrix, where $N$ is
the number of points of a $d$ dimensional parameter space.


The Saltelli sequence is obtained as follows: first we generate an \(N\) by
\(2d\) Sobol' sequence, (as demonstrated for the case of  $N=4$, $d=3$ in the
first line of  \cref{eqn:3_sobol}). Let the first \(d\) columns be called
submatrix \(\A\) (\cB{blue}), and the last \(d\) submatrix \(\B\) (\cR{red}).
The values in the matrices indicate the locations in parameter space at which
the model is to be evaluated, for parameters which can take on values over the
range 0 to 1. We next  construct a number \(d\) of \(N\) by \(d\) matrices
\(\AB\), for \(i\in\{1,2,...,d\}\), such that for each \(\AB\) the
\(i^{\mathrm{th}}\) column is taken from matrix \(\B\), while the remaining
columns come from matrix \(\A\) .  The matrices \(\A\), \(\B\) and \(\AB\)
specify all the points of the parameter space at which the model is to be
evaluated (one point per row), giving a total of $N\times(2+d)$ evaluations
which are required to calculate the first order sensitivity indices.

\begin{align*}
\mathrm{Sobol}(4,3)&=
\begin{bmatrix}
\cB{0.500} & \cB{0.500} & \cB{0.500} & \cR{0.500} & \cR{0.500} & \cR{0.500} \\
\cB{0.250} & \cB{0.750} & \cB{0.250} & \cR{0.750} & \cR{0.250} & \cR{0.750} \\
\cB{0.750} & \cB{0.250} & \cB{0.750} & \cR{0.250} & \cR{0.750} & \cR{0.250} \\
\cB{0.125} & \cB{0.625} & \cB{0.875} & \cR{0.875} & \cR{0.625} & \cR{0.125} \\
\end{bmatrix}\addtocounter{equation}{1}\tag{\theequation}\label{eqn:3_sobol}
\\
\mathbf{A}_\mathbf{B}^{\left(1\right)}&=
\begin{bmatrix}
\cR{0.500} & \cB{0.500} & \cB{0.500} \\
\cR{0.750} & \cB{0.750} & \cB{0.250} \\
\cR{0.250} & \cB{0.250} & \cB{0.750} \\
\cR{0.875} & \cB{0.625} & \cB{0.875} \\
\end{bmatrix}
\\
\mathbf{A}_\mathbf{B}^{\left(2\right)}&=
\begin{bmatrix}
\cB{0.500} & \cR{0.500} & \cB{0.500} \\
\cB{0.250} & \cR{0.250} & \cB{0.250} \\
\cB{0.750} & \cR{0.750} & \cB{0.750} \\
\cB{0.125} & \cR{0.625} & \cB{0.875} \\
\end{bmatrix}
\\
\mathbf{A}_\mathbf{B}^{\left(3\right)}&=
\begin{bmatrix}
\cB{0.500} & \cB{0.500} & \cR{0.500} \\
\cB{0.250} & \cB{0.750} & \cR{0.750} \\
\cB{0.750} & \cB{0.250} & \cR{0.250} \\
\cB{0.125} & \cB{0.625} & \cR{0.125} \\
\end{bmatrix}
\end{align*}

A visual impression of the different sampling approaches is given by
\cref{fig:3_sample} which shows five commonly used types of sampling: OAT,
uniform pseudo-random number generator, uniform grid sampling, a two-dimensional
Sobol' sequence and Saltelli sampling. The OAT approach is often used with far
fewer evaluations than shown here, which makes it computationally cheaper than
the other approaches. The drawback of this method is clear from the vast areas
of the parameter space that are left unexplored. This problem is only
exacerbated on  increasing the dimensionality of the parameter space. The
pseudo-random number generation suffers from poor convergence, as randomness
often results in over and under sampling of many regions. The Sobol' and
Saltelli sequences uniformly sample the parameter space and achieve the low
discrepancy target at a reasonable computational cost.

\subsection{Sensitivity indices}\label{sec:3_ind}

Given a scalar model $Y$ with independent inputs, we can define the first order
effect of the variance in the input $X_i$ as:
\begin{align}\label{eqn:3_var_i}
E_i&=E_{\Xwoi}\left(Y|X_i\right) = \int Y(X_i) {\rm pdf}(X_i) \prod_{i \neq j}^{d} {\rm d}X_i \\
V_i&=\Var_{X_i}\left(E_i\right) =  \int \left(E_i - E(Y)\right)^2 {\rm pdf}(X_i) {\rm d}X_i,
\end{align}
where $X_i$ is $i^{\mathrm{th}}$ model input, $V_i$ is the variance integrated
in $X_i$ space over dimension $i$, and $E_i$ is the mean $Y$ value, integrated
over the $d$-dimensional $X$ space in all dimensions except $i$.  Since $V_i$
can only take values between $0$ and $\Var(Y)$, the total variance in the model
output, we can define normalised first-order sensitivity indices $S_i$ of each
input parameter as
\begin{equation}\label{eqn:3_s_i}
S_i=\frac{V_i}{\Var\left(Y\right)},
\end{equation}
which measures the effect that varying the input $X_i$ has on the output,
averaged over variations of all other inputs.  If $S_i=1$, all variance in $Y$
comes from the variance in $X_i$, whereas if $S_i=0$, none of it does, and $Y$
is independent of $X_i$.

In order to measure the interactions between model parameters, we can define
higher order indices.  For second order interactions, the combined variance is
\begin{equation}\label{eqn:3_var_ij}
V_{ij}=\Var_{X_{ij}}\left(E_{\mathbf{X}_{\sim{ij}}}\left(Y|X_i,X_j\right)\right)-V_{i}-V_{j},
\end{equation}
from which $S_{i,j}$ can be calculated analogously to $S_i$.

It should now be obvious from the definition of the model variance why the OAT
methods are inappropriate for complex models -- they do not consider the full
contribution to the model variance given by \cref{eqn:3_var_i} (which averages
over {\it all} values of the other inputs, instead of being measured only at a
designated slice, as shown in the relevant panel of \cref{fig:3_sample}), nor
does OAT treat the combined variance of two (\cref{eqn:3_var_ij}) or more
variables correctly.

For a deterministic model, the only source of variance in the output is the
variances of the inputs.  Therefore, from variance decomposition it follows that
\begin{equation}\label{eqn:3_var}
\sum_{i=1}^{d}\Var_{i}+\sum_{i<j}^{d}\Var_{ij}+...+\Var_{12...d}=\Var(Y),
\end{equation}
which we normalise to obtain the sensitivity indices of all orders
\begin{equation}
\sum_{i=1}^{d}S_{i}+\sum_{i<j}^{d}S_{ij}+...+S_{12...d}=1.
\end{equation}

A direct consequence is that, in order to analytically decompose the total
variance of the model, one needs to compute variances of $2^{d}-1$ variables,
which can be computationally expensive for complex models.  However, if we
assume that the indices decrease as their order increases (which is correct for
the model of interest here), we might be less interested in the precise values
of higher-order contributions, and focus instead on the total higher-order
response of a given variable.  In this case it is convenient to combine the
higher-order terms into a total-order index
\begin{equation}\label{eqn:3_s_ti}
S_{Ti}=\frac{E_{\Xwoi}\left(\Var_{X_i}\left(Y|\Xwoi\right)\right)}{\Var(Y)}=1-\frac{\Var_{\Xwoi}\left(E_{X_i}\left(Y|\Xwoi\right)\right)}{\Var(Y)},
\end{equation}
containing all terms of the decomposed output variance which include $X_i$.
Unlike the first-order indices, the $S_{Ti}$ do not have to add up to $1$, as
they include all the input interactions\footnote{In this case, the whole is
literally more than the sum of the parts.}.

Higher order effects can also be calculated in simpler analyses, such as
Analysis of Variance \citep[ANOVA]{fisher1918}, High Dimensional Model
Representations \citep[HDMR]{sobol1993} or derivative-based methods.  However,
the total indices are a unique feature of the variance-based SA, and are a major
advantage of this methodology, as they allow for a direct comparison of the
linear and non-linear impacts of the input parameters.

Following \citet{jansen1999}, \citet{sobol2001} and \citet{saltelli2010}, we can
use the approximate forms of the first and total order sensitivity indices,
based on the sampling
matrices $\A,\B,\AB$.
\begin{align}\label{eqn:3_approx}
\Var_{X_i}\left(E_{\Xwoi}\left(Y|X_i\right)\right)&\approx\frac{1}{N}\sum_{j=1}^{N}f\left(\B\right)_{j}\left(f\left(\AB\right)_j-f\left(\A\right)_j\right)\\
E_{\Xwoi}\left(\Var_{X_i}\left(Y|{\Xwoi}\right)\right)&\approx\frac{1}{2N}\sum_{j=1}^{N}\left(f\left(\A\right)_j-f\left(\AB\right)_j\right)^{2},
\end{align}
where $f\left(\mathbf{X}\right)$ is the model $f$ evaluated at point $X$.

\begin{figure}
	\includegraphics[trim=0 0 0 0, width=1.0\columnwidth,height=1.0\columnwidth]{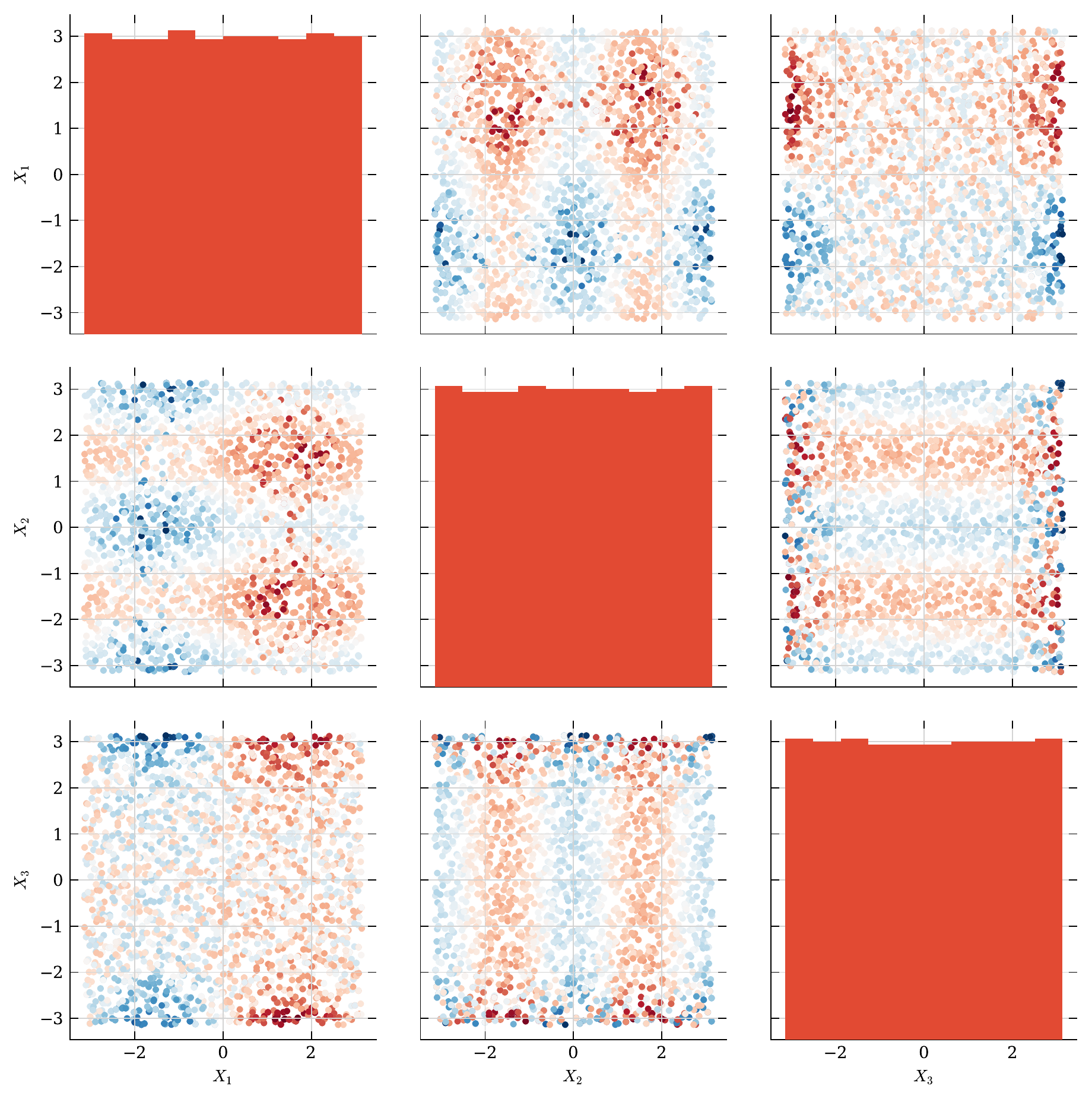}
	\caption[Ishigami function]{The 4000 evaluations of the  Ishigami function
	(\protect\cref{eqn:3_ishigami}.  On-diagonal histograms show distribution of
	the $X_i$ parameters.  Off-diagonal scatter plots show pairs of parameters
	and  are colour-coded to show the value of output $Y$.}
	\label{fig:3_ishigami}
\end{figure}

\begin{figure}
	\includegraphics[trim=30 20 0 10, width=1.1\columnwidth,height=1.1\columnwidth]{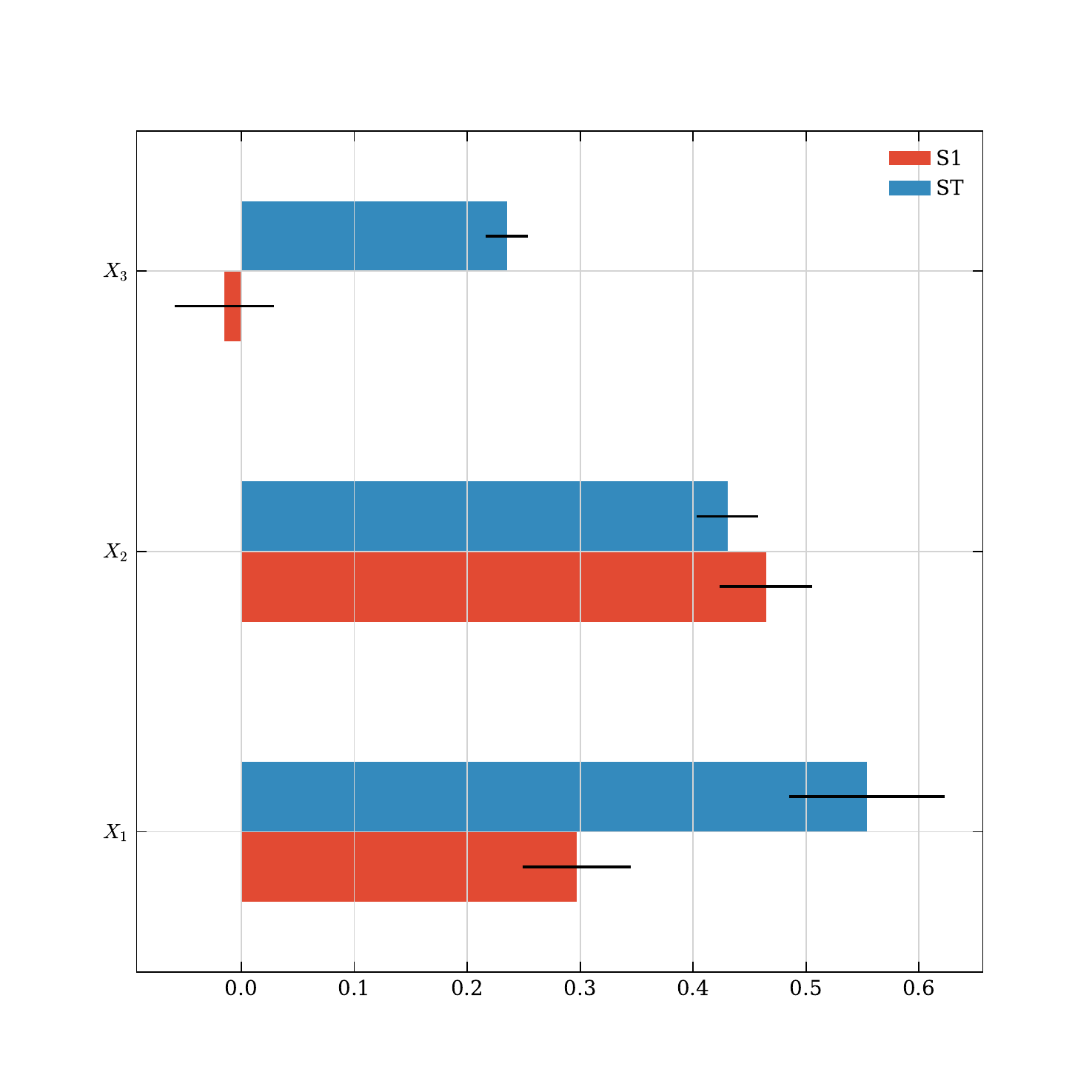}
	\caption[Sensitivity Indices of the Ishigami function]{First-order (red) and
	total (blue) sensitivity indices of the three input parameters of the
	Ishigami function, with $1\sigma$ confidence bars (black).}
	\label{fig:3_ishigami_sa}
\end{figure}

\begin{figure}
	\includegraphics[width=\columnwidth,height=\columnwidth]{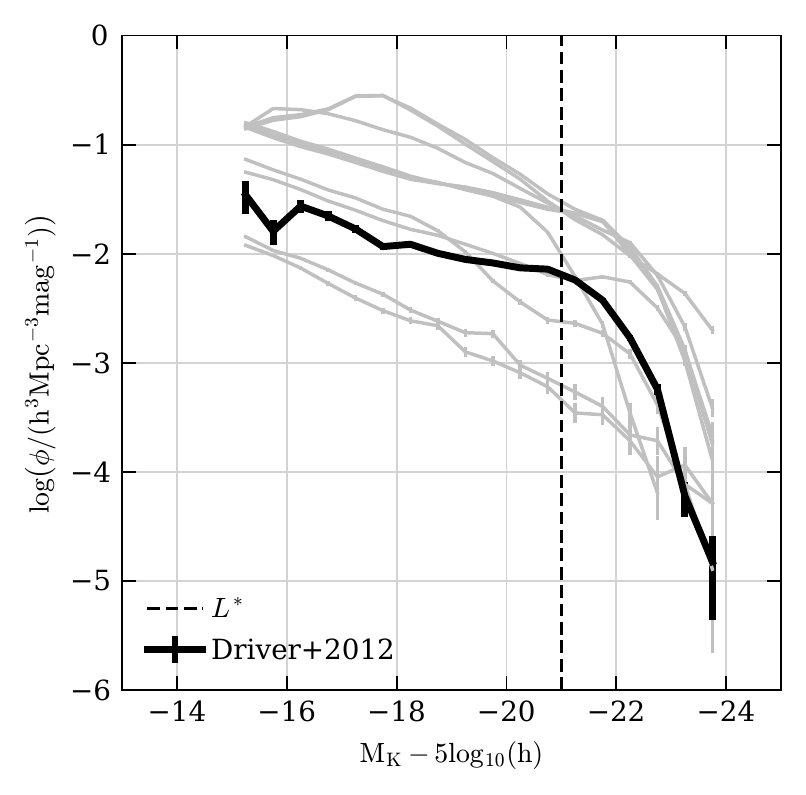}
	\caption[$K$-band luminosity function]{The $K$-band LF at $z=0$ in the AB
	magnitude system.  Gray lines represent 10 \GF model realisations randomly
	chosen from the 1600-model run series.  The black line represents the
	observational data from \protect\citet{driver2012}.  The black vertical line
	is drawn at $L=L^*$, and separates the bright and the faint ends of the LF.}
	\label{fig:3_lfk}
\end{figure}

\begin{figure*}
\includegraphics[width=\textwidth,height=0.5\textwidth]{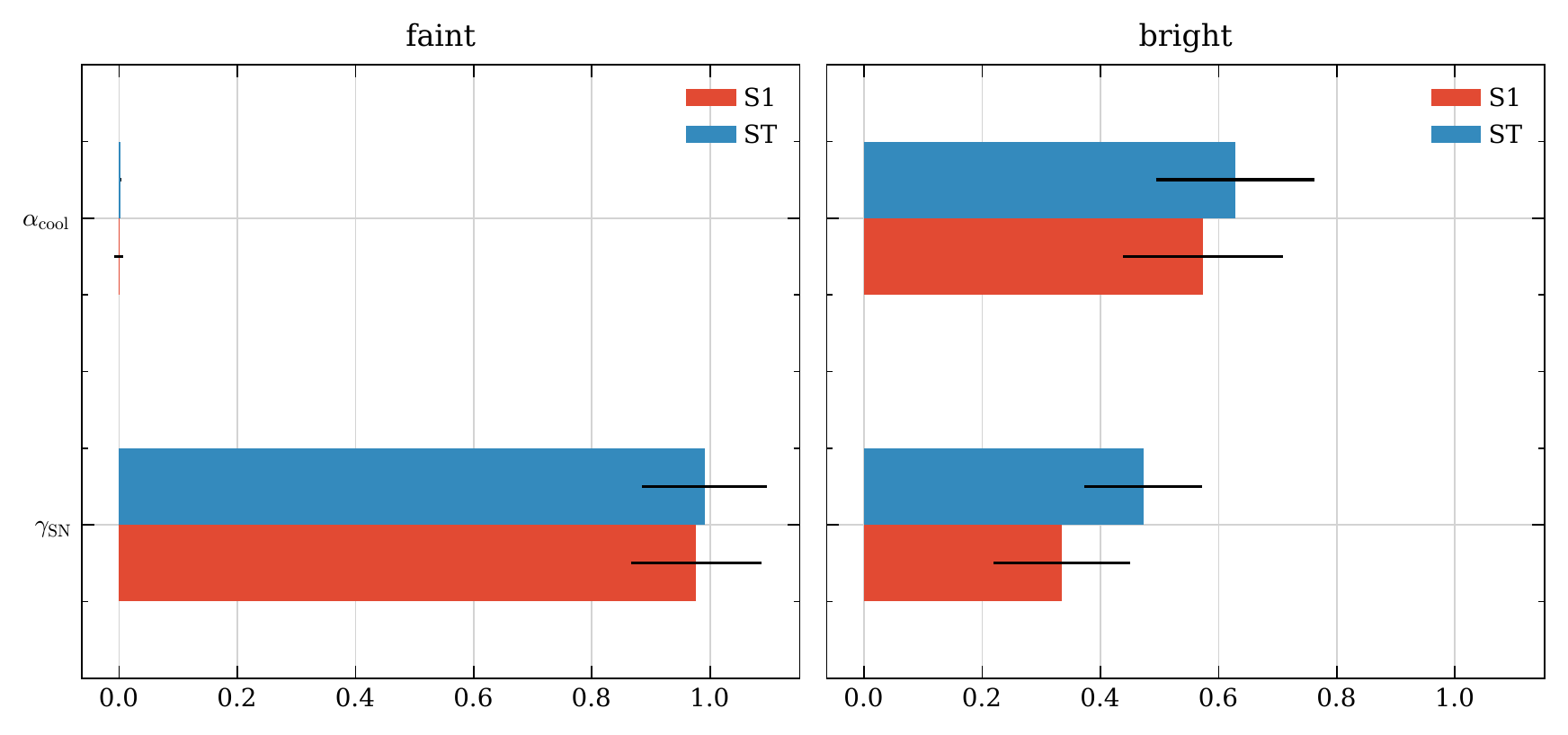}
\caption[Sensitivity indices for the case in which the LF is measured coarsely
using two bins, on allowing two parameters to vary: $\alpha_{\rm cool}$ and
$\gamma_{\rm SN}$, varying 2 parameters at a time]{Sensitivity indices for the
first series of runs, when varying two parameters in \GF\!:
$\alpha_{\mathrm{cool}}$ and $\gamma_{\mathrm{SN}}$, for the $K$-band luminosity
function measured in two coarse luminosity bins. The colours of the bars
indicate different indices, first (blue)  (\protect\cref{eqn:3_s_i}) and total
(red) (\protect\cref{eqn:3_s_ti}) order for a given variable.  The left panel
shows indices for $L<L^*$, and the left for $L>L^*$ (see
\protect\cref{eqn:3_out_2bin}). The black bars show the $1\sigma$ confidence
interval for the sensitivity indices.}
\label{fig:3_sa_2bin_2param}
\end{figure*}

\begin{figure}
\includegraphics[width=\columnwidth,height=\columnwidth]{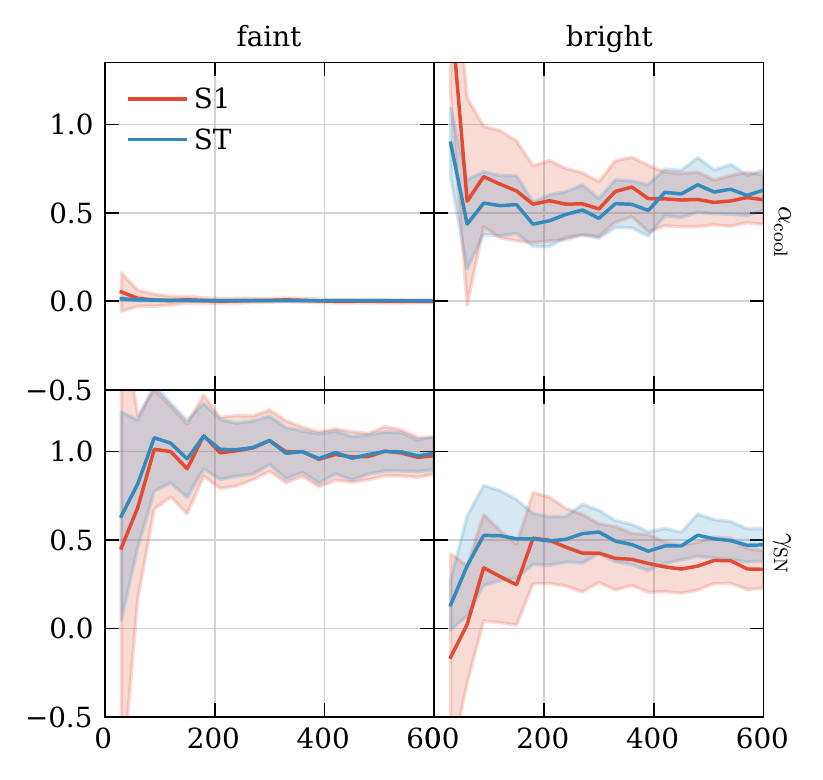}
\caption[]{Convergence of the first- and total-order sensitivity indices for the
first series of runs, when varying 2 parameters of \GF\!,
$\alpha_{\mathrm{cool}}$ and $\gamma_{\mathrm{SN}}$, as a function of a number
of samples.  Individual subplots correspond to the faint and bright end of the
$K$-band LF (columns, labelled on the top), and the $\alpha$ parameters (rows,
labelled on the right).  Solid lines correspond to the values of the indices,
and the shaded regions to the $1\sigma$ confidence band of the values, both
colour-coded by the order of the indices as labelled in the legend.}
\label{fig:3_conv}
\end{figure}

\subsection{Illustrative sensitivity analysis of a toy model}\label{sec:3_ishigami}

The performance of the sensitivity analysis estimator can be demonstrated using
a toy model.  The Ishigami function is an example of such a model, and is
commonly used to test the predictions of sensitivity analysis because it
contains non-linear interacting terms. Nevertheless, the sensitivity indices can
be calculated analytically and compared to the estimated values.

The Ishigami function is defined by Eq.~14  of \citet{ishigami1991} as
\begin{equation}\label{eqn:3_ishigami}
Y\left(X_1,X_2,X_3\right)=\sin\left(X_1\right)+a\sin^2\left(X_2\right)+b\,X_3^4\,\sin\left(X_1\right)
\end{equation}
where the $X_i$ are random variables uniformly distributed between $-\pi$ and
$\pi$, such that ${\rm pdf}(X_i) = U(-\pi, \pi)$, and $a,b$ are numerical
constants, here chosen to be $7$ and $0.1$ respectively.


The SA was carried out by running the model on inputs generated by a
3-dimensional Sobol' sequence for 500 realisations, which resulted in $4000 =
500 \times (2 + 2 \times 3)$ values of  $X_i$  (as explained in
\cref{sec:3_saltelli}). Next, \cref{eqn:3_ishigami} was evaluated at each $X_i$
point, giving a vector $Y$ of length 4000.   Finally, the vector $Y$ was
analysed using the \texttt{SALib} Python package \citep{herman2017}.

The evaluations of \cref{eqn:3_ishigami} are shown in \cref{fig:3_ishigami}, and
the first- and total- order sensitivity indices of the three input parameters
are shown in \cref{fig:3_ishigami_sa}.  It is interesting to draw some
qualitative observations from \cref{fig:3_ishigami}:
\begin{itemize}
	\item  varying $X_1$ and $X_2$ in isolation results in large changes in $Y$;
	this is reflected by large values for $S_{1}$ and $S_{2}$.

	\item  varying $X_1$ and $X_2$ together has a large effect on $Y$; this is reflected by large values for $S_{T1}$ and $S_{T2}$.

	\item  varying $X_3$ for mid-range values produces little effect, but
	varying other parameters at extreme $X_3$ values produces a large change in
	$Y$; correspondingly, $S_3$ is nearly zero, but $S_{T3}$, which captures the 
	global response of $Y$ to $X_3$, is larger
\end{itemize}

A more complete SA would involve computing second order indices, and comparing
sensitivity indices for different versions of the Ishigami function, such as
with different $a$, $b$ parameters, or over  different $X_i$ ranges.  However,
this more complete analysis is beyond the scope of this section, as it is only
meant for demonstration purposes.  The source code used to reproduce this
analysis has been made public:
\url{https://github.com/oleskiewicz/sensitivity/releases/tag/v1.0}.

\subsection{\GF output used in the sensitivity analysis}\label{sec:3_out}

When applying SA to a model with a multi-dimensional output, it is necessary to
select the most interesting outputs manually.  The Sobol' index method assumes,
and can only be calculated for, separate one-dimensional output vectors
$\mathrm{Y}$. From the formal standpoint this is problematic as the sensitivity
indices contain no information about any correlations between various model
outputs.  However, in practice one could perform model runs which follow the
Saltelli sampling and then carry out separate sensitivity analyses for any
desired number of model outputs, since running the model is more time consuming
than calculating the Sobol' indices.

Here we focus on the prediction of \GF for the $K$-band luminosity function (LF)
at $z=0$, calculated as described in \cref{sec:3_lf}.  We have chosen to
consider this statistic due to the well-understood influence of the model
parameters on the form of the luminosity function (see the extensive discussion
in \citealt{lacey2015}).  Varying the parameters around the values used in the
fiducial model shows that the bright and faint ends of the luminosity function
are regulated by different physical processes.  Therefore, the sensitivity
indices could be easily verified for errors, and we will be able to quantify our
intuition regarding the relative importance of the different feedback modes on
the abundance of galaxies at different luminosities.

We have elected to perform the analysis on the model output values normalised by
the observational data (see \cref{eqn:3_out}) instead of on the model output
itself.  This way, the values we focused on were close to the ones typically
used for model optimisation, and had a reduced dynamic range, being effectively
normalised by the observational values.  Analysing a SAM independently of the
observational constraints, while interesting in its own merit, is outside the
scope of this work.

\section{Results}\label{sec:3_results}

\begin{figure*}
\includegraphics[trim=0 270 0 0,clip,width=\textwidth]{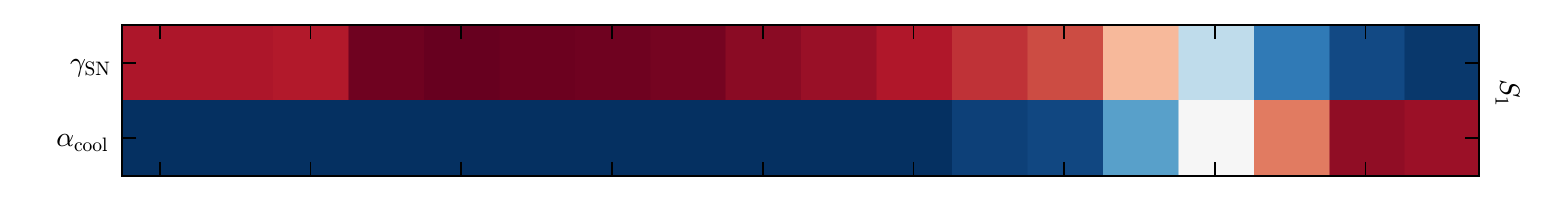}
\includegraphics[height=0.8\textwidth,width=\textwidth]{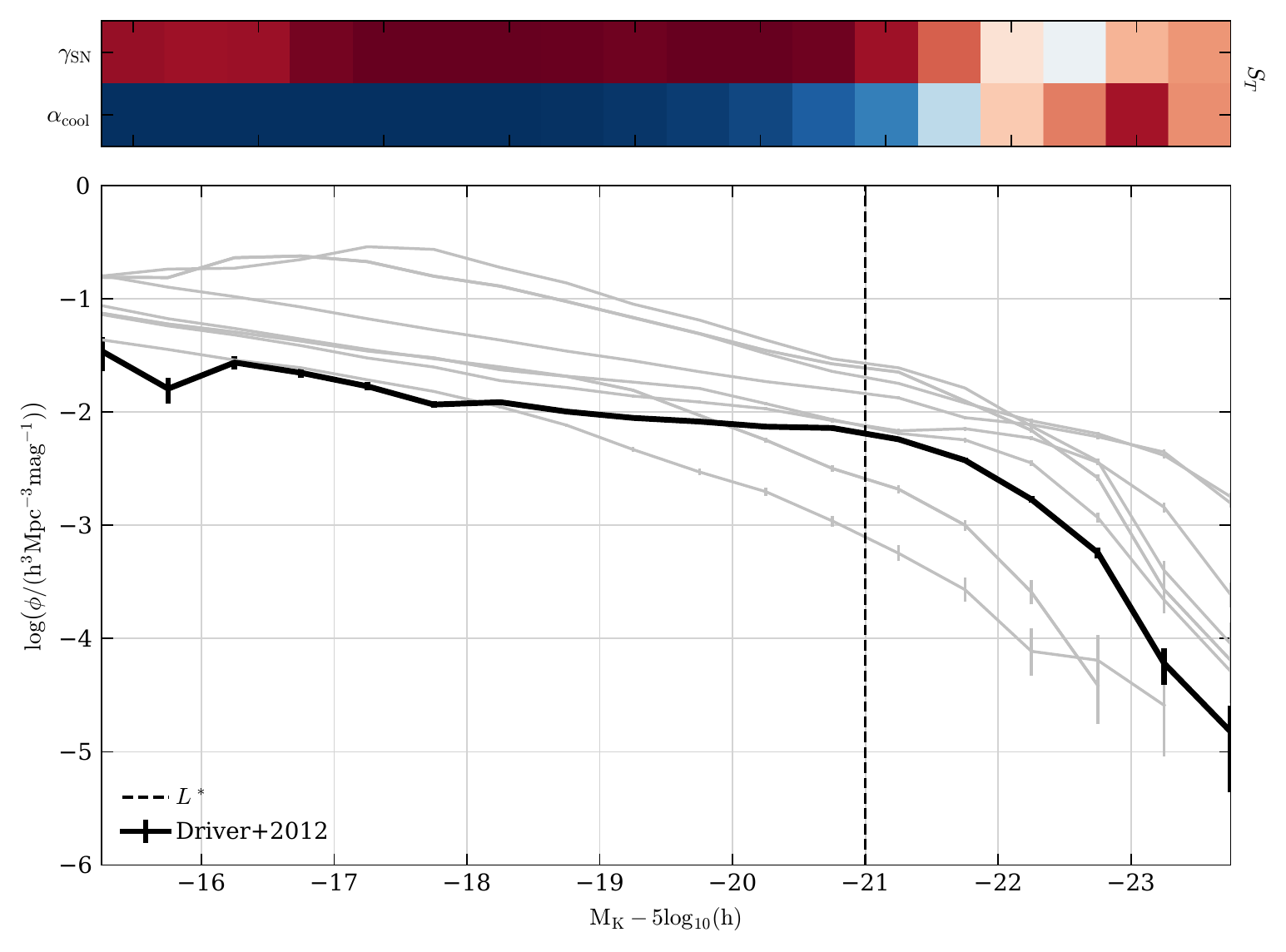}
\caption[First-order sensitivity indices for all bins computed after varying 2
parameters at a time]{First- and total-order sensitivity indices
(\protect\cref{eqn:3_s_i}) for the first series of runs, when varying 2
parameters of \GF\!: $\alpha_{\mathrm{cool}}$ and $\gamma_{\mathrm{SN}}$.  {\it
Bottom panel:} $K$-band luminosity function at $z=0$ like in
\protect\cref{fig:3_lfk}; gray lines correspond to 10 randomly chosen runs;
black line is the observational data from \protect\citet{driver2012}; dashed
vertical line corresponds to $L^*$.  {\it Top panels:} first- and total-order
(as labelled on the right) sensitivity indices of two variables (y axis) for
individual magnitude bins (x axis), colour-coded by value between $0$ (not
sensitive, dark blue) and $1$ (most sensitive, dark red).}
\label{fig:3_sa_2param}
\end{figure*}

\begin{figure*}
\includegraphics[width=\textwidth]{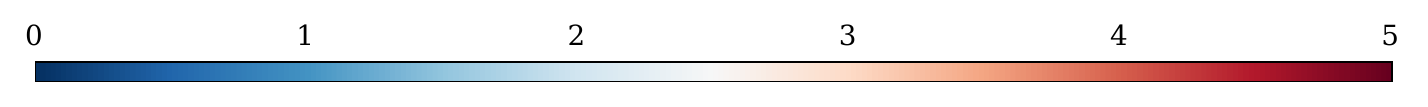}
\includegraphics[width=\textwidth,height=\textwidth]{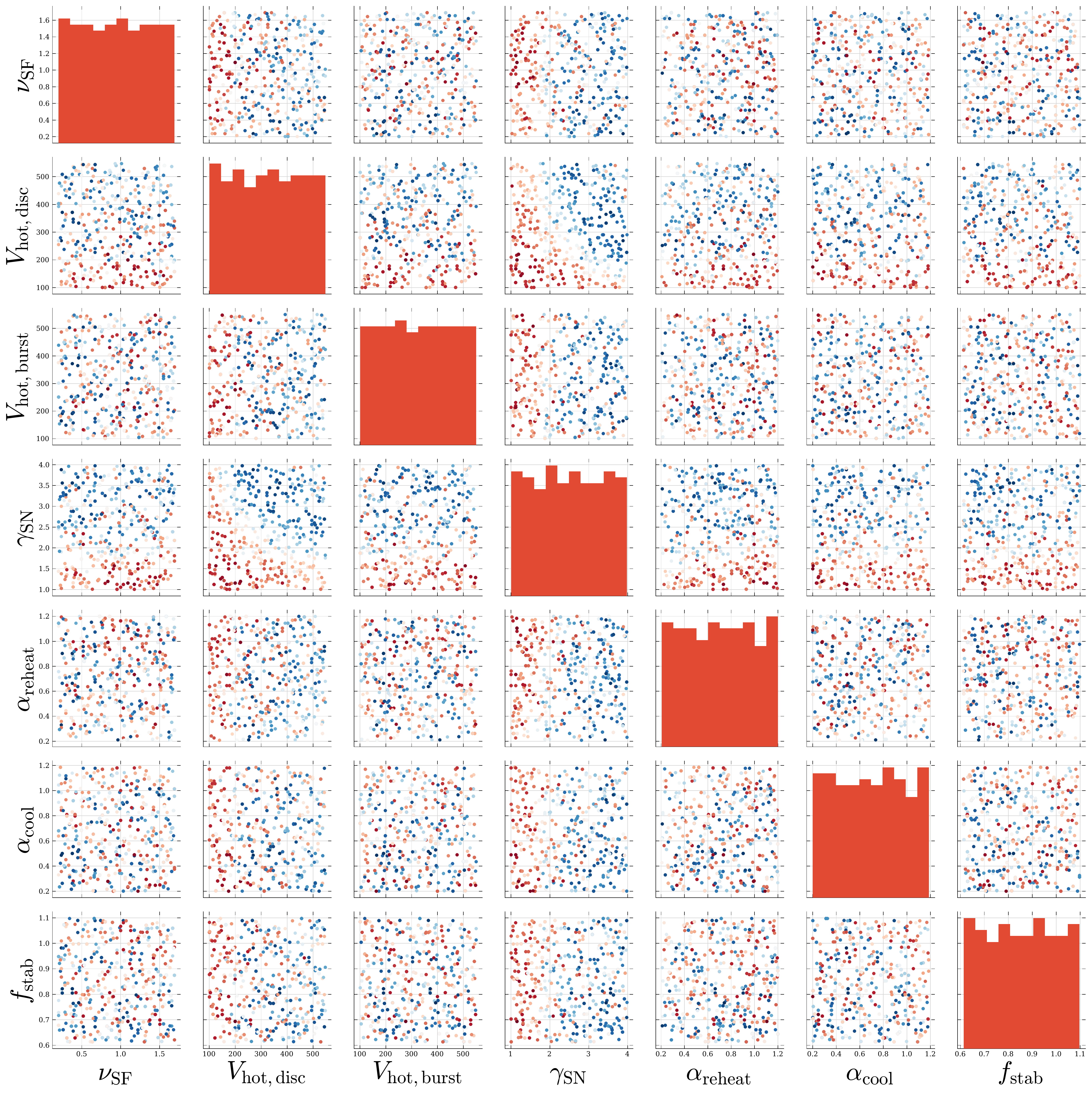}
\caption[$\chi^2$ of \GF models for the luminosity function in $K$-band]{The
parameter space of the second \GF experiment in which 7 parameters are varied
over 1600 realisations of the  model.  On-diagonal histograms show the
nearly-uniform distribution of the individual parameters, as expected for
Saltelli sampling.  Off-diagonal scatter plots show the parameter space for
pairs of parameters,  colour-coded by the goodness-of-fit
(\protect\cref{eqn:3_chi2}) of the model prediction for the $K$-band luminosity
function (\protect\cref{fig:3_lfk}), to the observational estimate from
\protect\citet{driver2012}; blue points correspond to runs with low values of
\protect\cref{eqn:3_out}, as labelled by the colourbar.}
\label{fig:3_chi2}
\end{figure*}

\subsection{Sensitivity analysis experiments}

We have carried out two separate sensitivity analyses using \GF: (1) 600 \GF
model runs varying two parameters ($\alpha_{\mathrm{cool}}$ and
$\gamma_{\mathrm{SN}}$), and (2) 1600 model \GF runs varying seven  parameters
(see \cref{tbl:3_param}).

For both series of runs, a SA was carried out on the $K$-band LF at $z=0$,
parametrised in two different ways.  

In the first instance we performed a simple analysis splitting the LF into two
broad luminosity bins, one covering a range of luminosities brighter than $L^*$
and the other luminosities fainter than $L^*$ (see \cref{fig:3_lfk}).  For each
run, we calculated two model outputs covering the bright and faint ends of the
LF, $d_{\mathrm{faint}}$ and $d_{\mathrm{bright}}$, defined by summing the
normalised differences between the observed and predicted values of the
luminosity function for luminosities brighter and fainter than  $L^*$. e.g.:
\begin{equation}\label{eqn:3_out_2bin}
d_{\mathrm{faint}}=\sum_{L<L^*}\frac{\log_{10}(\phi)-\log_{10}(\hat{\phi})}{\log_{10}(\phi)},
\end{equation}
with $d_{\mathrm{bright}}$ defined analogously for $L>L^*$.  The observed
luminosity function $\hat{\phi}$ is taken from \citet{driver2012}.  Unlike a
traditional measure of model fitness, we {\em do not} take the absolute value or
square of the distance between the model prediction and the data.  This is
because the sign of the output (i.e. the sense of the discrepancy) is valuable
information for the sensitivity indices, as it contains the direction of the
model response.

This coarse analysis is quantitatively identical to measuring the LF using only
two broad luminosity bins. This exercise has two goals: (1) to verify that SA
produces explainable results which can be interpreted in accordance with our
physical intuition about the galaxy formation model, and (2) to check the
convergence of the sensitivity indices and their confidence intervals, which can
be estimated as explained in \cref{sec:3_sens}.

After this coarse two-bin analysis, in the second case we calculate sensitivity
indices for each of the 18 luminosity bins, $L_{i}$,  using the quantity:
\begin{equation}\label{eqn:3_out}
d_{\mathrm{i}}=\frac{\log_{10}(\phi_{i})-\log_{10}(\hat{\phi_{i}})}{\log_{10}(\phi_{i})}.
\end{equation}
This serves as a fine-grained analysis, which can quantify the relative impact
of different parameters on the individual segments of the LF, as well as
uncovering interactions between model parameters.

\subsection{Feedback processes and the luminosity function}\label{sec:3_2param}

The first series of runs, which analysed the effects of changing two of the
parameters which specify different feedback processes in \GF, $\alpha_{\rm
cool}$ and $\gamma_{\rm SN}$, was carried out to verify the usefulness of the SA
and to evaluate its effectiveness, given our physical intuition, regarding the
expected impact on the LF of varying these model parameters. Only two parameters
were allowed to vary to speed-up the analysis and allow for an easier
interpretation of results: $\alpha_{\mathrm{cool}}$ and $\gamma_{\mathrm{SN}}$
(see \cref{tbl:3_param} for the range of parameter values considered).  Recall
that $\gamma_{\rm SN}$ controls the mass loading of SNe driven winds and
$\alpha_{\rm cool}$ determines the halo mass above which AGN heating shuts down
the cooling flow. 

\cref{fig:3_sa_2bin_2param} shows the first- and total-order sensitivity indices
calculated from 600 \GF model runs for the coarse-bin analysis of luminosity
function using two bins, one fainter and one brighter than $L^*$.  The results
are striking, but match our expectations: it is clear that $\gamma_{\rm SN}$ is
the dominant parameter out of the two in shaping the model output for galaxies
fainter than $L^*$ (and hence, that such galaxies are mainly affected by SNe
feedback) and that both parameters have similar significance for galaxies
brighter than $L^*$ (albeit $\alpha_{\rm cool}$ is slightly more important), and
so bright galaxies are affected by SNe feedback and AGN heating.  Moreover,
${\rm S}_1$ and ${\rm S}_{\rm T}$ are comparable in all cases, which means that
the model response to varying these parameters is mostly linear.

\cref{fig:3_conv} shows the convergence of the indices from the
\cref{fig:3_sa_2bin_2param} as a function of the number of samples $N$. The
indices do not change substantial after 100 \GF runs. 

\cref{fig:3_sa_2param} shows the first- and total-order sensitivity indices for
the fine-grained analysis of the LF using multiple luminosity bins.  We can see
that $L^*$ is close to coinciding with the bin at which AGN heating starts to
become important, which explains the results shown in
\cref{fig:3_sa_2bin_2param}.  We also learn that while SNe feedback does not
interact with the AGN heating at the faint end of the LF, their influence
over the bright end is strongly correlated.

We did not consider the best-fit model for this two parameter case, since we
perform a rudimentary estimate of the best-fitting  parameter set in the next
section, when varying more \GF model parameters at the same time.

\subsection{Sensitivity analysis over a multi-dimensional parameter space}\label{sec:3_7param}

\begin{table}
	\centering
	\caption{The best-fit \GF parameters found in this work, in relation to
	\protect\citet{driver2012}.}
	\label{tbl:3_fit}
	\begin{tabular}
		{lr}
		\toprule
		parameter                                   & value  \\
		\midrule
		$\nu_{\mathrm{SF}}$ [\si{\per\giga\yr}]     & 0.46   \\
		$\gamma_{\mathrm{SN}}$                      & 3.45   \\
		$\alpha_{\mathrm{reheat}}$                  & 0.74   \\
		$V_{\mathrm{hot,disc}}$ [\si{\km\per\sec}]  & 332.69 \\
		$V_{\mathrm{hot,burst}}$ [\si{\km\per\sec}] & 392.90 \\
		$\alpha_{\mathrm{cool}}$                    & 0.58   \\
		$f_{\mathrm{stab}}$                         & 0.77   \\
		\bottomrule
	\end{tabular}
\end{table}

The design of the second experiment, in which seven \GF parameters are varied
simultaneously (\cref{tbl:3_param}), is inspired by the work on parameter
optimisation using Bayesian emulators by \cite{bower2010} and
\cite{rodrigues2017}. For comparison, we use the same parameter ranges adopted
in their studies. This exercise requires significantly more model realisations
than the first one, since we sample a higher dimensional parameter space and aim
to observe interactions between more parameters.

\cref{fig:3_chi2} shows the parameter space and its sampling, colour-coded by
the goodness-of-fit measure
\begin{equation}\label{eqn:3_chi2}
\chi^2=\sum_i \frac{\left(\log_{10}(\phi_i)-\log_{10}(\hat{\phi_i})\right)^2}{\log_{10}(\phi_i)},
\end{equation}
where the sum is carried out over all luminosity bins and low values of $\chi^2$
are blue.  While $\chi^2$ is not a robust model output for SA, as it does not
contain information about the {\it direction} of the model response as explained
in \cref{sec:3_results}, it is still a useful measure of a global model response
or ``quality of fit''. The on-diagonal histograms indicate that the Saltelli
sampling produces a nearly uniform sampling of parameter space, as expected from
a low-discrepancy sequence.  The off-diagonal scatter plots give a first
indication of some of the first-order index results: the $\chi^2$ of the model
LFs is sensitive to variation of $\gamma_{\rm SN}$, is degenerate in the
$\gamma_{\rm SN}$--$v{\rm hot,disc}$ plane (which follows directly from
\cref{eqn:3_beta}), and depends only weakly on other parameters.

\begin{figure*}
\includegraphics[width=\textwidth,height=0.5\textwidth]{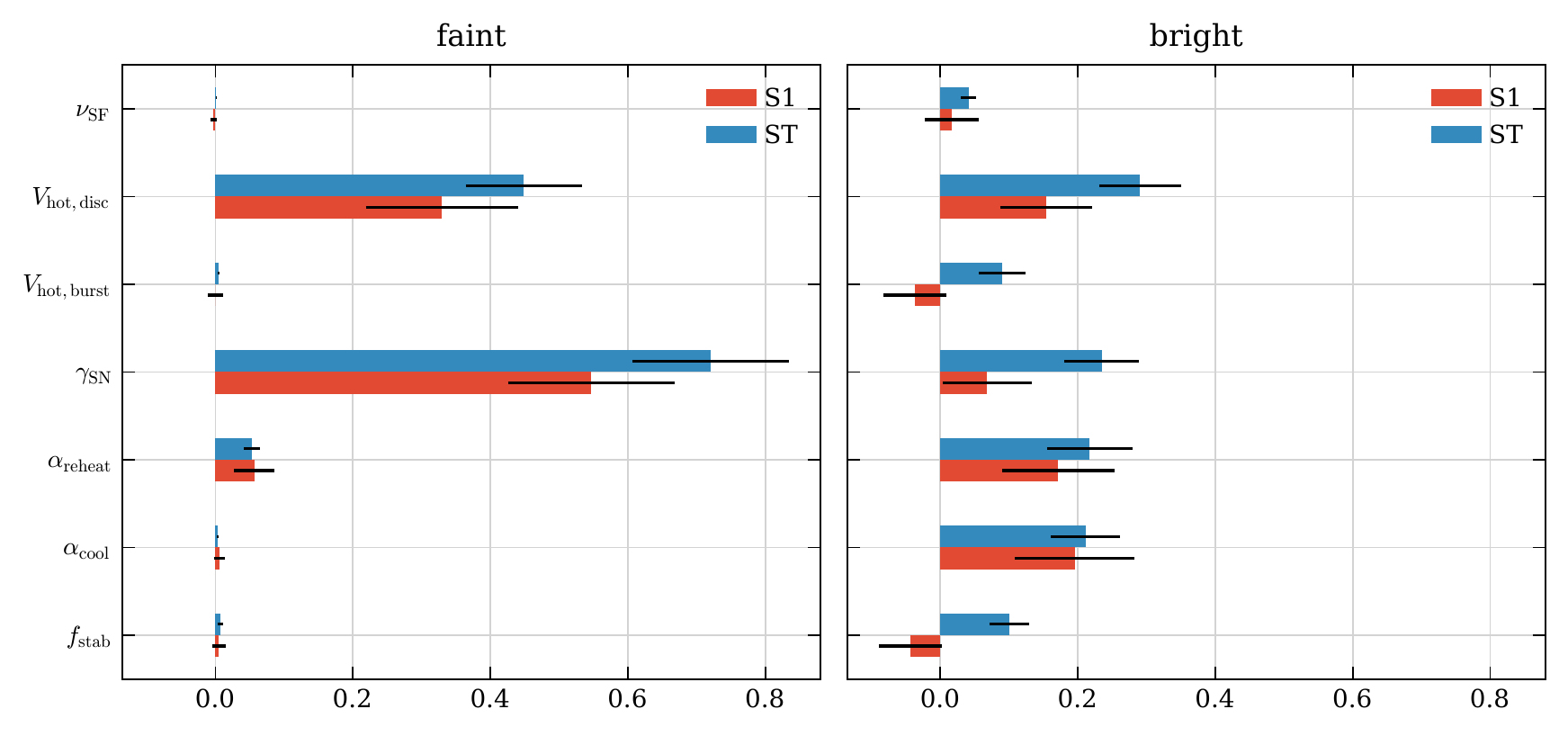}
\caption[Sensitivity indices for 2 bins computed after varying 7 parameters at a
time]{Sensitivity indices for the second series of 1600 \GF runs, varying 7
model parameters (\protect\cref{tbl:3_param}), computed using the coarse two bin
description of the luminosity function. The bar colours indicate the values of
different indices, the first index (\protect\cref{eqn:3_s_i}) (S1, red) and
total order index  (\protect\cref{eqn:3_s_ti}) (ST, blue) for each parameter.
The left panel shows indices for galaxies in the luminosity bin fainter than
$L^*$, and the right panel for galaxies in the bin brighter than  $L^*$
(\protect\cref{eqn:3_out_2bin}). The black bars show the $1\sigma$ confidence
intervals for the sensitivity indices.}
\label{fig:3_sa_2bin_7param}
\end{figure*}

\begin{figure*}
\includegraphics[trim=0 240 0 0,clip,width=\textwidth]{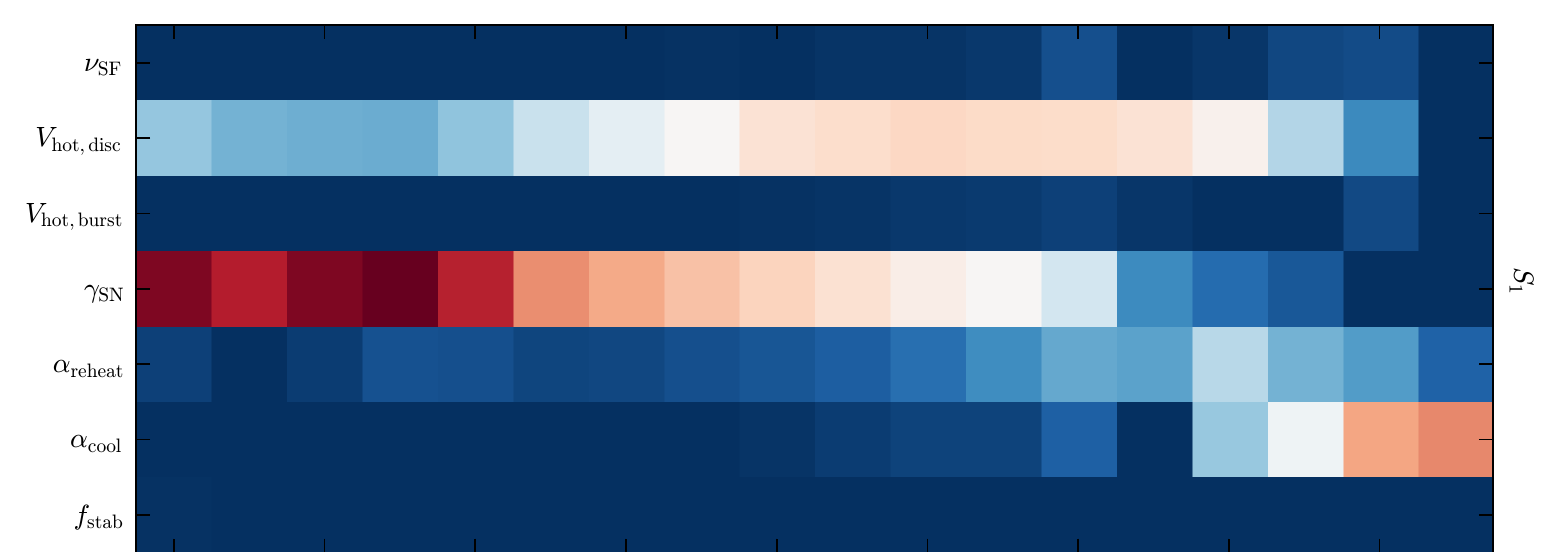}
\includegraphics[height=0.8\textwidth,width=\textwidth]{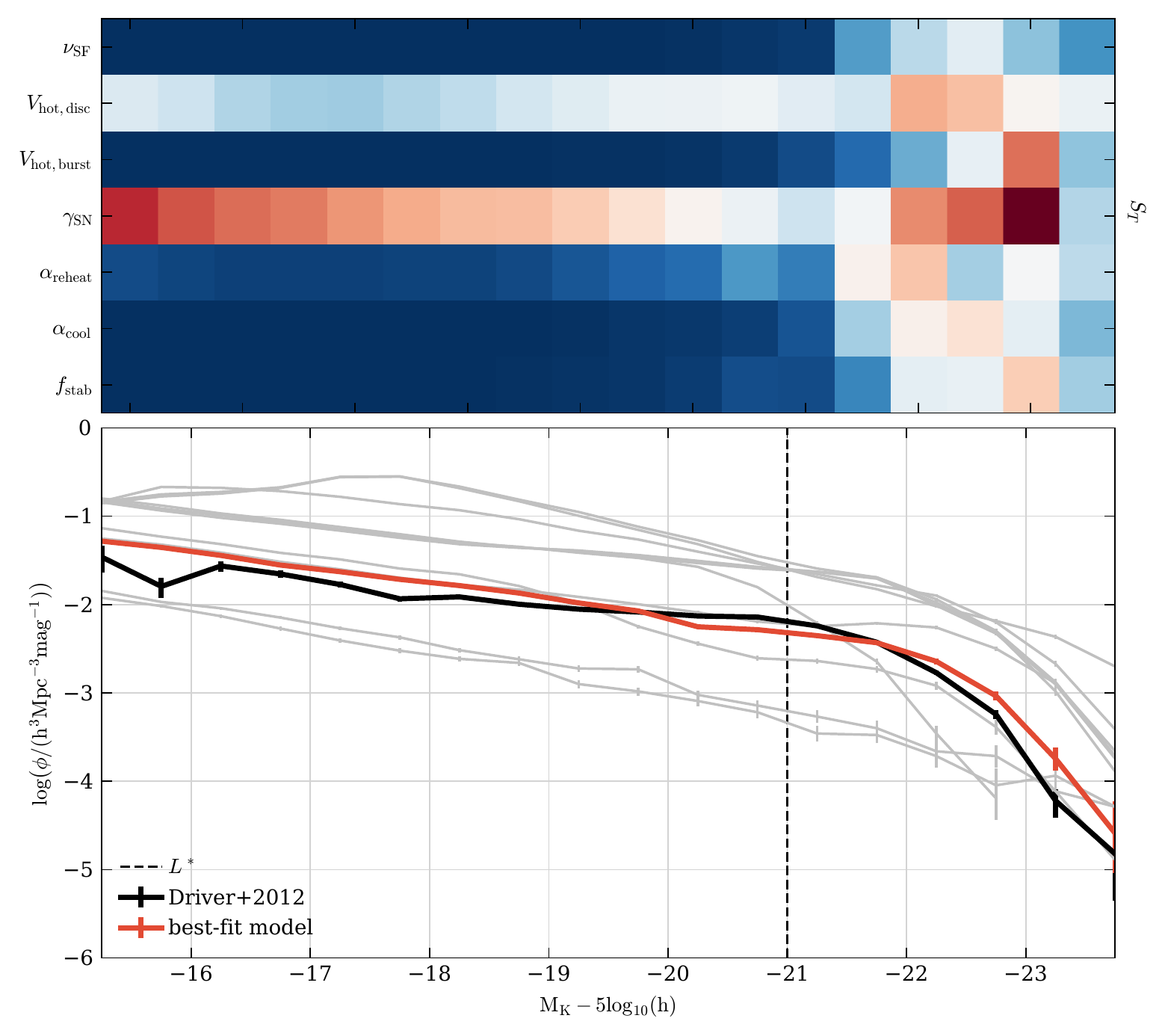}
\caption[First- and total-order sensitivity indices for all bins computed after
varying 7 parameters at a time]{First- and total-order sensitivity indices
(\protect\cref{eqn:3_s_i}) for the second series of runs, when varying 7 \GF
parameters (\protect\cref{tbl:3_param}). {\it Bottom panel}: $K$-band luminosity
function at $z=0$ as in \protect\cref{fig:3_lfk}; grey lines show 10 randomly
chosen \GF models; the black line connects observational data from
\protect\citet{driver2012}; dashed vertical line corresponds to $L^*$; the solid
red line shows the best-fitting model.  {\it Top panels:} first- and total-order
(as labelled on the right) sensitivity indices of two variables (y axis) for
individual magnitude bins (x axis), colour-coded by value between $0$ (not
sensitive, dark blue) and $1$ (most sensitive, dark red).}
\label{fig:3_sa_7param}
\end{figure*}


\cref{fig:3_sa_2bin_7param} shows the first- and total-order sensitivity indices
for the coarse-binned analysis of the LF.  Since sensitivity indices are derived
from the normalised variance (\cref{eqn:3_s_i,eqn:3_s_ti}), the values should
always be between 0 and 1, which they are after allowing for the confidence
interval.  Similarly to \cref{fig:3_sa_2bin_2param}, in
\cref{fig:3_sa_2bin_7param} we see two different types of behaviour of the \GF
model: the faint end is dominated by SNe feedback, while the bright end has a
mixed, non-linear response to many parameters. Interestingly, while AGN feedback
(via $\alpha_{\rm cool}$) has the highest first-order sensitivity index (S1) for
the bright end of the LF, the total-order indices (ST) of SNe feedback processes
dominate.  Of particular interest are the  $f_{\rm stab}$ and $v_{\rm
hot,burst}$ parameters. These parameters have nearly zero first-order response
indices (which means that their impact cannot be detected by an OAT analysis),
but their combined higher-order responses are significant.

It is instructive to see the origin of the values reported in
\cref{fig:3_sa_2bin_7param}, by inspecting how the sensitivity changes
bin-by-bin in \cref{fig:3_sa_7param}. The results are consistent with
\cref{sec:3_2param}, and together provide an interpretation of the behaviour of
the \GF model.  Moreover, displaying the model output together with model
sensitivity can be of use when manually tweaking the model, allowing for a fine,
manual control over the precise details of the LF (or, indeed, other outputs).

Finally, we note that \cref{fig:3_sa_7param} also shows the LF for the
best-fitting model, as determined by the smallest value of \cref{eqn:3_chi2}.
This can be considered an additional benefit of running SA -- requiring so many
model realisations naturally finds one which is likely to be close to a global
optimum.  The best-fitting  parameter values are reported in \cref{tbl:3_fit}.
Note that the values diverge from those reported in \citet{lacey2015}, due to
different fitting method and the fact that this study only considered the
$K$-band LF, whereas \citet{lacey2015} took into account multiple observations
in a manual parameter tuning.  Of particular interest is the value of $V_{\rm
hot,disc}$, which is over $20\%$ larger than in the previous calibration of this
\GF model.  We attribute this difference to the fact that, as discussed in
\cref{sec:3_7param} and shown in \cref{fig:3_sa_2bin_7param}, the combined
total-order sensitivity index of $V_{\rm hot,disc}$ outweighs the first-order
index for both ends of the $K$-band LF. This suggests that the optimal value of
this parameter could be missed by OAT model fitting.  The differences in the
other parameter values are not as significant as they might seem -- the
variables with the highest sensitivity match the previously reported values
pretty closely (e.g. $\gamma_{\mathrm{SN}}$ is within $7\%$), and the variables
with low sensitivity that diverge by a significant margin by definition of the
sensitivity indices do not have significant impact on the $K$-band LF.

\section{Conclusions}\label{sec:3_conclusions}

We have used variance-based sensitivity analysis to analyse the sensitivity of
the $K$-band luminosity function predicted using the \GF semi-analytical model
of galaxy formation to the variation of the model  parameters.  We have shown
that sensitivity analysis is a useful tool, which goes beyond simple model
fitting and one-at-a-time parameter variation, and we have demonstrated that it
can be applied to a challenging problem in computational astrophysics.
Variance-based sensitivity analysis is perhaps particularly useful for the
semi-analytic modelling of galaxy formation modelling, due to the computational
expense of searching a multi-dimensional parameter space and the non-linearity
of the model. These features have led some to view such models as  black boxes.
Part of the aim of the sensitivity analysis presented here is to make the
behaviour of the model and how it responds to parameter changes more
transparent.  

In its present form sensitivity analysis can only deal with one-dimensional
outputs of a model, which on the one hand means that it cannot be used to
resolve correlations in model outputs (such as between the predictions in
different bins of the luminosity function or between the luminosity function in
different bands; see \citealt{benson2014}), yet on the other hand this feature
gives the scientist performing the study unlimited flexibility in choosing and
parametrising the outputs they find the most important. Here, we have elected to
perform the sensitivity analysis using the model predictions in luminosity bins
cast in terms of the difference between the computed and measured $K$-band
luminosity function at $z=0$. Our motivation for this was that by choosing an
established observable with a well understood connection to the underlying
physical processes and their description in term of \GF parameters, we could
make a convincing case for the usefulness of the sensitivity analysis.

With this in mind, future work on SA might want to examine the variance of the
outputs of the semi-analytic model alone, independently of the corresponding
measured observable values.  There are three main reasons for such an approach:
i) using the full dynamic range of the predictions:  normalising the model
output by observations flattens the dynamic range, and while SA works equally
well for small and large values, by only analysing a flat version of the model
predictions we effectively take the regions in which the model gives a flat or
steep response (for instance, the faint and bright end of the LF respectively)
and make them look the same.  ii) independence of post-processing: by comparing
to data, we had to make a choice about the norm of the discrepancy between the
model output and observations: do we retain the sense of the discrepancy or
square it?  A different SA study could have chosen differently, altering the
results.  By analysing model outputs independently of the observations these
choices are no longer necessary. iii) data independence: SA results could change
if a different dataset is used with the same model.

Moreover, the $K$-band luminosity function is just one possible output and there
are many others which a successful semi-analytic model should reproduce
accurately. Analysing all of these is outside the scope of this study, but we
hope to have shown that SA is a promising avenue of research.

Finally, we note that while correctly estimating model sensitivity can be useful
in guiding model optimisation and improving the physical interpretation of the
parameters of the galaxy formation models, one must remember that even the most
rigorous sensitivity analysis can only provide the answers with regards to the
model, not the underlying physical system itself \citep{taleb2013}. Therefore,
the relationship between the structure of the model and that of the physical
system remains open to discussion.